\documentclass[aps,pra,twocolumn,showpacs,superscriptaddress,nobibnotes,10pt]{revtex4-1}
\usepackage{amsmath,graphicx,amssymb}
\usepackage{xcolor}
\usepackage{soul}
\begin{document}

\title{Anderson localization of light in dimension $d-1$}

\author{Carlos E Maximo}
\affiliation{Departamento de F\'{\i}sica, Universidade Federal de S\~{a}o Carlos, Rod. Washington Lu\'{\i}s, km 235 - SP-310, 13565-905 S\~{a}o Carlos, SP, Brazil}
\author{Noel A Moreira}
\affiliation{Instituto de F\'{i}sica de S\~{a}o Carlos, Universidade de S\~{a}o Paulo - 13560-970 S\~{a}o Carlos, SP, Brazil}
\affiliation{Universit\'{e} C\^{o}te d'Azur, CNRS, INPHYNI, France}
\author{Robin Kaiser}
\affiliation{Universit\'{e} C\^{o}te d'Azur, CNRS, INPHYNI, France}
\author{Romain Bachelard}
\affiliation{Departamento de F\'{\i}sica, Universidade Federal de S\~{a}o Carlos, Rod. Washington Lu\'{\i}s, km 235 - SP-310, 13565-905 S\~{a}o Carlos, SP, Brazil}
\affiliation{Universit\'{e} C\^{o}te d'Azur, CNRS, INPHYNI, France}
%
%

\date{\today}



\begin{abstract}
Localization of electromagnetic waves in disordered potentials is prevented by polarization terms, so only light scattering systems of dimensions $d=1$ and $2$ with scalar properties exhibit light localization. We here discuss the presence of surface modes in vectorial systems of dimensions $d=2$ and $3$, which possess lower-dimensional scattering properties, and present features of Anderson localization. In particular, vectorial waves in $3D$ ($2D$) presents surface localized modes with scaling consistent with localization $2D$ ($1D$).
\end{abstract}

\maketitle
\section{Introduction}
Introduced in the context of electronic transport by P.W. Anderson~\cite{Anderson1958}, the localization of waves in a disordered potential was later understood to be a general wave phenomenon. Acoustic~\cite{He1986,Hu2008}, matter~\cite{Billy2008,Roati2008,Kondov2011,Semeghini2015}, surface plasmon polaritons~\cite{Shi2018} or electromagnetic waves~\cite{John1983,John1987, Genack1991,Wiersma1997, Schuurmans1999,Strzer2006,Garcia2012,Sperling2013} were shown to localize as well, and the universal ``scaling analysis''~\cite{Abrahams1979}, which does not account for the microscopic details, suggested that the only critical parameter was the dimension: one- and two-dimensional systems from the orthogonal symmetry class  always present localization features, being hindered only by the finite system size. In contrast, three-dimensional systems localize above a critical disorder.

A shadow was cast on this universal picture when it was discovered that the vectorial character of eletromagnetic waves prevents localization~\cite{Skipetrov2014,Bellando2014}. Initially discussed for three-dimensional systems, for which the conclusions of experimental reports had been already questioned~\cite{Wiersma1997,Schuurmans1999,Scheffold2013,Sperling2016}, a similar result was reported for  two-dimensional light scattering ~\cite{Maximo2015}: In-plane polarizations, which are coupled to each other, present vectorial properties and do not localize, whereas the polarization orthogonal to the plane, uncoupled from the others, behaves as a scalar wave and exhibits localized modes~\cite{Laurent2007}. This situation now calls for an experimental verification of the role of polarization, as it was for example proposed in Ref.\cite{Skipetrov2015} by imposing a strong magnetic field to decouple the different scattering channels.

From a theoretical point of view, the scaling theory of localization, that focuses on the thermodynamic limit, has been recognized as the standard method to demonstrate the transition to localization in any type physical systems. Nevertheless, it does not take into account finite-size phenomena that arise on the boundaries of the samples. In this paper, we show that the presence of surface modes in homogeneous samples leads to the emergence of Anderson-localized modes at the surface of vectorial-waves systems.  More specifically, vectorial electromagnetic waves in dimension $d=2$ and $3$ present scattering modes with exponentially-decaying profiles along the surface, with localization lengths that scales as the mean-free path in dimension $d=1$ and $2$. 

Studies of Anderson localization of light waves usually consider distributions prone to surface modes, i.e., with homogeneous densities (to yield homogeneous localization properties) and high densities (to reach strongly scattering samples). Anderson localization {\it a priori} refers to the thermodynamics limit of infinite samples, where the proportion of surface modes vanishes. Nevertheless, these localized surface modes for vectorial waves may still be very relevant since on the one hand, the samples used to study Anderson localization are usually rather small in terms of optical wavelengths, and on the other hand they correspond to the boundary between the scattering medium and the surrounding one.

In the next section, we present the microscopic model used to describe the light scattering in dimension $2$ and $3$. In Section \ref{sec:2d}, we present our analysis, as well as the localization at the surface of $2D$ systems. In Section \ref{sec:3d}, these results are extended to $3D$ systems. Finally, in Section \ref{sec:discussion} we draw our conclusions and discuss the perspective for surface localization.

\section{Point-scatterer model}
To address the vectorial light scattering problem in dimensions $d=2$ and $3$, we use a microscopic model that describes $N$ point-like two-level dipoles, with positions $\mathbf{r}_j$, coupled through the vacuum modes of the $d$-dimensional space. In the linear optics regime, the formal Markovian integration of the vacuum modes dynamics yields the following effective interactions between pairs of dipoles~\cite{Lehmberg1970,Maximo2015,Hill2017}:
\begin{eqnarray}
K_{\alpha\beta}^{2D}\left(\mathbf{r}_{jl}\right) & = & \delta_{jl}\delta_{\alpha\beta}+\left(1-\delta_{jl}\right)\big[\delta_{\alpha\beta}H_{0}\left(kr_{jl}\right)\label{eq:2d}
\\ &&+\left(1-\delta_{\alpha\beta}\right)e^{2\alpha i\phi_{jl}}H_{2}\left(kr_{jl}\right)\big], \nonumber
\\ K_{\alpha\beta}^{3D}\left(\mathbf{r}_{jl}\right) & = & \delta_{jl}\delta_{\alpha\beta}+\left(1-\delta_{jl}\right)\big[\delta_{\alpha\beta}f\left(k r_{jl}\right)\label{eq:3d}
\\ && +\mathcal{D}_{\alpha\beta}\left(\hat{\mathbf{r}}_{jl}\right)g\left(k r_{jl}\right)\big],\nonumber
\end{eqnarray}
where the polarization components are $\alpha,\beta=\pm1$ for 2D and $\alpha,\beta=0,\pm1$ for 3D.  $H_n$ is the Hankel function of the first kind and of order $n$, and $r_{jl}=\|\mathbf{r}_{jl}\|$ the Euclidean distance in a given dimension, with $\mathbf{r}_{jl} \equiv \mathbf{r}_j-\mathbf{r}_l$. Furthermore, $\hat{\mathbf{r}}_{jl}=\mathbf{r}_{jl}/r_{jl}$ represents the unitary vector
\begin{align}
\hat{\mathbf{r}}_{jl}=\begin{cases}
\left(\cos\varphi_{jl},\sin\varphi_{jl}\right) & \text{ for } d=2,\\
\left(\sin\theta_{jl}\cos\varphi_{jl},\sin\theta_{jl}\sin\varphi_{jl},\cos\theta_{jl}\right) & \text{ for } d=3.
\end{cases}
\end{align}
Finally, we have introduced the functions:
\begin{eqnarray}
f\left(x\right) & =& \frac{3}{2}\frac{e^{ix}}{ix}\left(1-\frac{1}{x^{2}}+\frac{i}{x}\right),\\
g\left(x\right) & =& \frac{3}{2}\frac{e^{ix}}{ix}\left(\frac{3}{x^{2}}-\frac{3i}{x}-1\right),\nonumber\\
\mathcal{D}_{\alpha\beta}\left(\hat{\mathbf{r}}_{jl}\right) & =& \left(-1\right)^{\delta_{\alpha\beta}+1}e^{i\left(\beta-\alpha\right)\varphi_{jl}}\cos^{2}\theta_{jl}\left(\frac{\tan\theta_{jl}}{\sqrt{2}}\right)^{\left|\alpha\right|+\left|\beta\right|}.\nonumber
\end{eqnarray}
The resulting collective scattering modes are given by the eigenvectors $\Phi$ of the non-Hermitian Green's matrices \eqref{eq:2d} and \eqref{eq:3d}~\cite{Skipetrov2014,Negro2019}. Note that in the present form, these matrices describe the resonant scattering by two-level systems, with a unitary linewidth and a resonant energy arbitrarily fixed to zero. Thus, the imaginary part of their eigenvalue associated to each mode corresponds to their shift in energy relative to this resonance, which can be selected by tuning appropriately the frequency of the driving light~\cite{Moreira2019}.

In two dimensions, the vectorial nature of the scattering manifests itself in Eq.\eqref{eq:2d} through the near-field term $H_2$, which is known to suppress exponentially localized modes over the sample area~\cite{Maximo2015}. Such term is absent in the two-dimensional scalar model, in which all the features of Anderson localization persist~\cite{Laurent2007,Garcia2012,Maximo2015}. Similarly, a scalar light approximation can be performed in three dimensions, where the polarization term $\mathcal{D}_{\alpha\beta}$ is neglected and only the leading order in $f$ for $r\gg 1$ is kept. While this scalar simplification induces a nonphysical transition to localization, the full 3D vectorial model predicts an absence of transition~\cite{Skipetrov2014, Bellando2014}. For this reason, throughout this work we focus on the vectorial models, which are not expected to present any localized modes.

We here consider random distributions $\mathbf{r}_j$ with homogeneous densities $\rho_d$. This corresponds for instance to the case of white paint samples~\cite{Strzer2006, Sperling2013} or macroporous GaP samples~\cite{Schuurmans1999}. This is also the approach adopted by theorists, as the localization properties depend on the disorder strength, which is here tuned through the scatterer density~\cite{Skipetrov2018}. Furthermore, the existence of a critical disorder threshold to reach the localization transition in 3D requires strongly-scattering samples. Considering an effective-medium point of view, highly-scattering and homogeneous samples naturally lead to the emergence of surface modes~\cite{Bachelard2012}.
\begin{figure}[!h]
\centering
\includegraphics[width=0.48\textwidth]{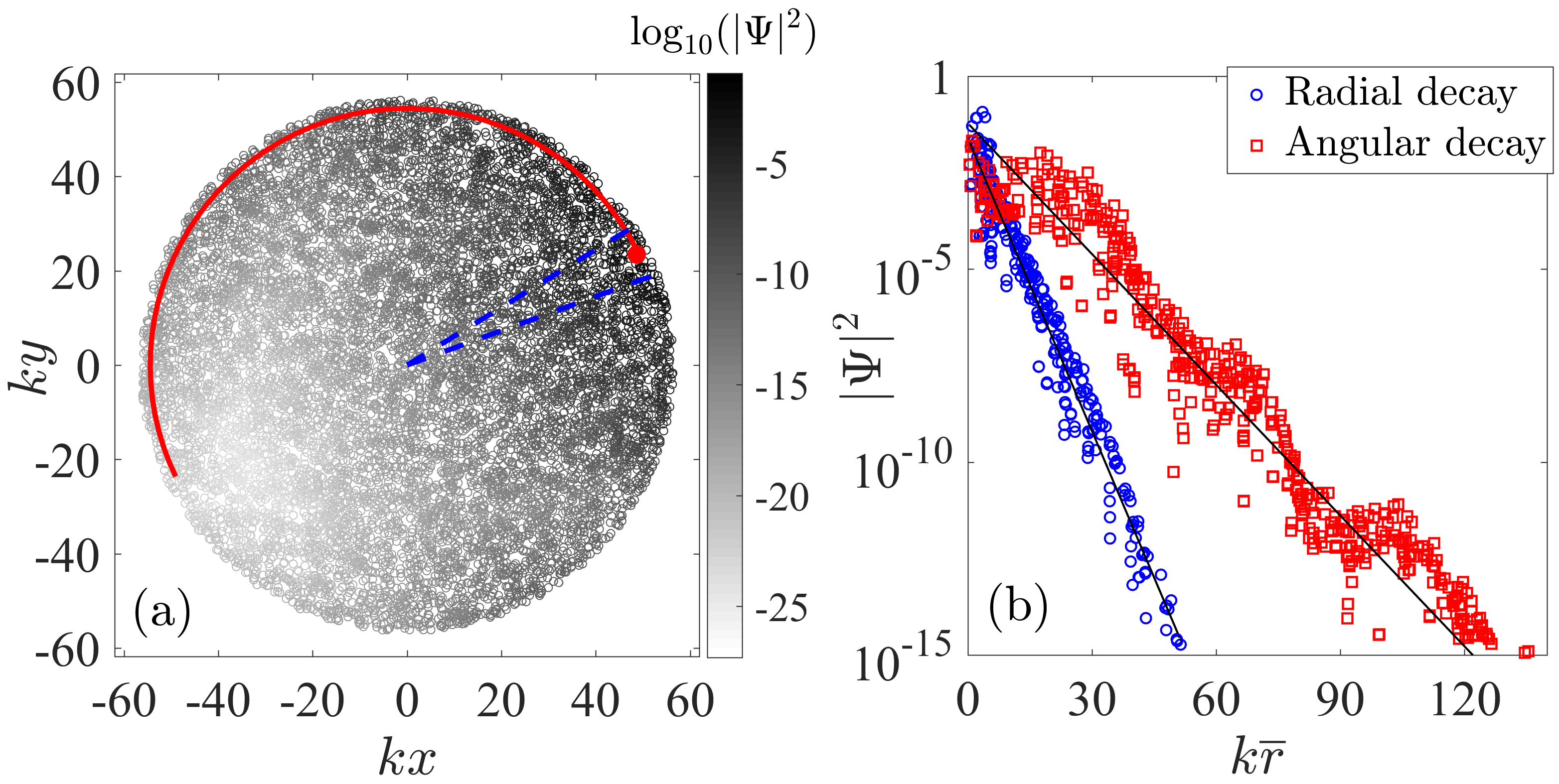} 
\caption{(a) 2D profile of a surface Anderson-localized mode. The red line starts at the center of mass, and depicts the thin layer where the angular profile is computed, whereas the blue dashed lines delimit the slice where the radial profile is computed. (b) Spatial decay along the radial axis and along the surface ('angular decay'), with the black lines corresponding to the exponential fit. $\overline{r}$ refers to the radius $r$ for the radial decay, and to $R\phi$, with $R$ the cloud radius. Simulations realized for $N=10^4$ scatterers and a density $\rho_2= k^2$.\label{fig:2dmode}}
\end{figure}

\section{Methods and surface localization in 2D\label{sec:2d}}

Once the disorder is accounted for, the surface modes, which possess the scattering properties of a medium of dimension $d-1$, are candidates for Anderson localization. This is illustrated in Fig.~\ref{fig:2dmode}, where a surface Anderson-localized mode from the 2D vectorial model \eqref{eq:2d} is presented. A closer inspection reveals in Fig.~\ref{fig:2dmode}(b) that there are two 'localization' mechanisms, as can be expected from the above discussion. On the one hand, the radial confinement, visible from the exponential radial decay of
\begin{equation}
|\Psi_{j}|^{2}=\sum_{\alpha=-1}^{+1}|\Phi_{j}^{\left(\alpha\right)}|^{2}, \label{eq:psi}
\end{equation}
originates in the high scatterer density; this mechanism is the same as for whispering-gallery modes, well described by Mie theory using the refractive index approach~\cite{vandeHulst1957}, and corresponds to a kind of Purcell effect, with increased scattering into the surface modes. On the other hand, Anderson localization manifests as an exponential decay of the mode amplitude along the surface [labelled 'angular decay' in Fig.~\ref{fig:2dmode}(b)], and is the result of the presence of disorder. Macroscopic oscillations can be observed in this angular decay profile, which we attribute to a resonance condition, where the whispering gallery mode must present an integer number of periods along the system boundary.

\begin{figure}[!h]
\centering
\includegraphics[width=0.485\textwidth]{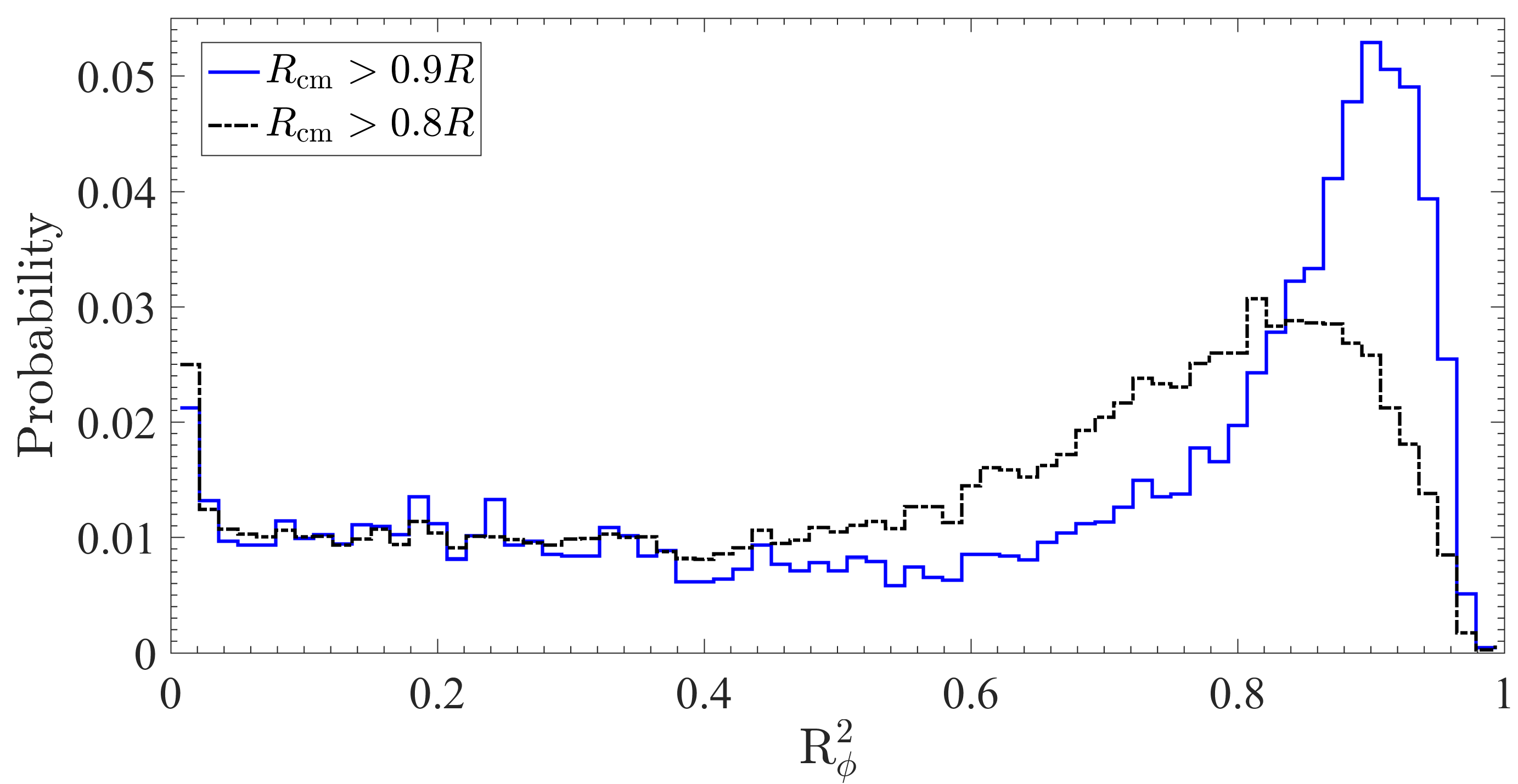} 
\caption{Histogram of the R$^2$ coefficient for the angular fit, using $70$ regular bins, and obtained for clouds with $\rho_2= k^2$, $N= 4000$, using $21$ realizations.\label{hist}}
\end{figure}

In order to identify surface modes, we first select the modes whose center of mass $|\mathbf{r}^{\text{cm}}|$ is at most at a distance $R/10$ from the surface, with $R$ the radius of the cloud. We observe that the closer to the surface, the more exponential-like is the decay of the mode profile.
A skin depth $\delta_m$ of the mode $m$ is then defined through the fitting $e^{-|\mathbf{r}_j-\mathbf{r}_m^{\text{cm}}|/\delta_m}$, performed over the atoms in a radial slice (see Fig1(a)). We have chosen to consider modes as surface mode when the exponential fit yields a confidence level above $0.9$ (using the coefficient of determination R$^2$). Then, we select the surface localized modes by performing an angular fitting $e^{-R\left(\phi_j-\phi^{\text{cm}}_m\right)/\xi_m}$ over the atoms in a layer $\delta_m$, only for the surface modes, with $\phi^{\text{cm}}$  the angular coordinate of the  center of mass and $\xi_m$  the angular localization length (see Fig1(b)). Using this protocol, the majority of surface modes (i.e., with an exponential radial decay from the surface) appear to be localized along the surface as well, with an exponential angular decay: This is illustrated in Fig.~\ref{hist}, where the histogram of the R$^2$ fitting coefficient, for the angular profile, presents a stronger peak close to unity for modes closer to the surface.

Let us now analyze qualitatively these confining mechanisms. In dimension $d$, an effective medium approach for an homogeneous sample of scatterer density $\rho_d$ predicts the presence of waves confined along the surface over a skin depth~\cite{Jackson98}
\begin{equation}
\delta_d\sim\Re(1/k\sqrt{n^2-1}), \label{eq:skindepth}
\end{equation}
with $n$ the refractive index. This results in the surface modes propagating in a $(d-1)$-dimensional disordered medium whose effective scatterer density and mean-free path are given by
\begin{subequations}
\begin{align}
\rho_{d-1}^\textrm{s}&=\rho_d \delta_d,
\\ l_{d-1}^\textrm{s}&=\frac{1}{\sigma_{d-1}\rho_{d-1}^\textrm{s}},
\end{align}
\end{subequations}
where $\sigma_{d}=8/k$, here refers to the scattering cross-section for scalar light in dimension $d=2$ ($\sigma_{d}=1$ for $d=1$).

\begin{figure}[!h]
\centering
\includegraphics[width=0.48\textwidth]{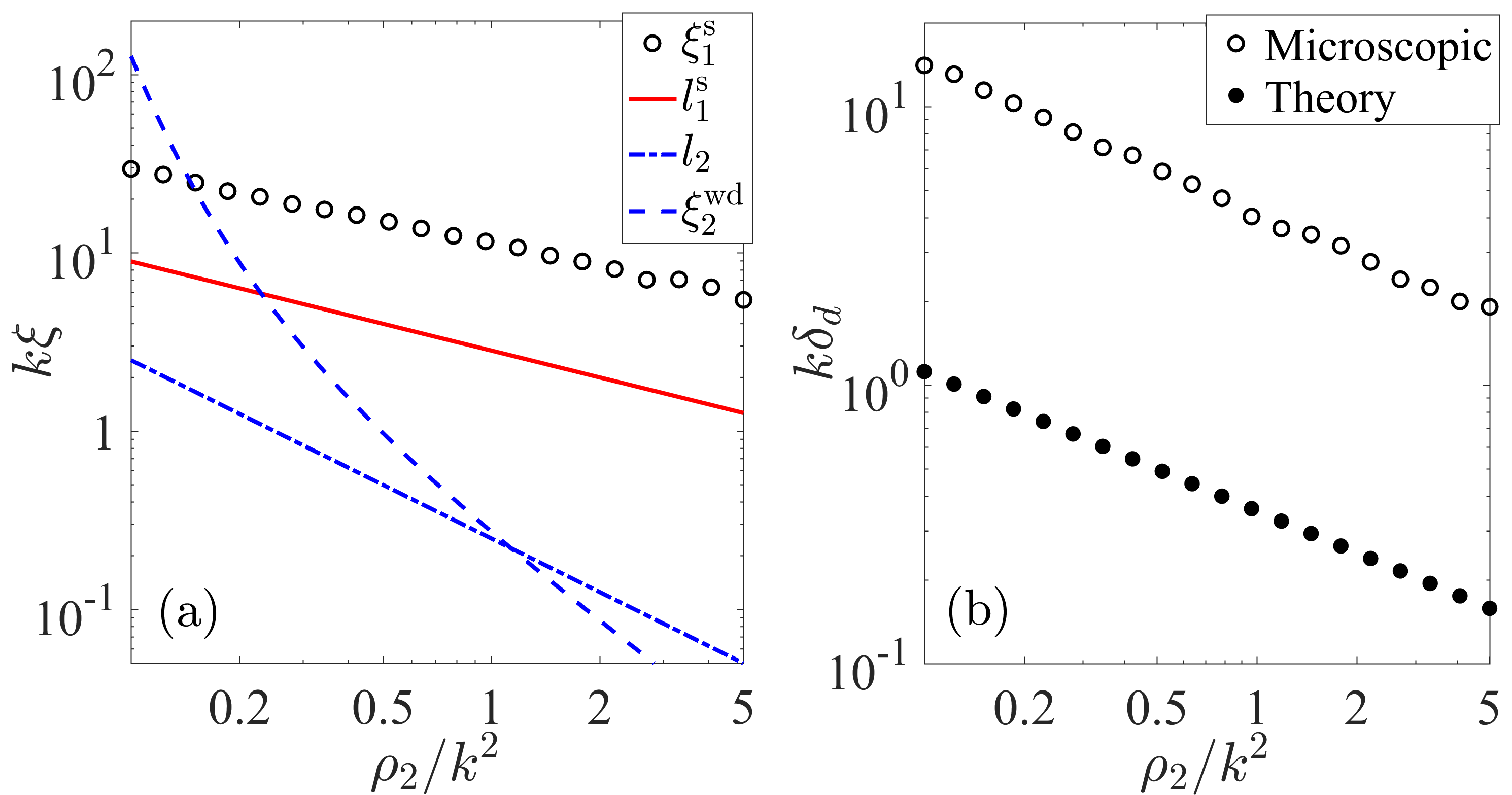} 
\caption{(a) Localization length $\xi_1^\text{s}$ and (b) skin depth $\delta_d$ of Anderson-localized surface modes, as a function of the sample density $\rho_2$, for the 2D vectorial scattering problem. It is compared to the 1D surface localization length $l_1^\textrm{s}$ and to the 2D mean free path $l_2$ and weak-disorder prediction $\xi_2^\textrm{wd}$ (see main text for details). Simulations realized for a cloud of $N=10^4$ scatterers, where $\xi$ is computed using an average over all localized surface modes, and for a single realization for each value of density.\label{fig:2d1d}}
\end{figure}

Let us first consider the two dimensional case. Surface modes then propagate in a $d=1$ space, where light is known to localize for any amount of disorder~\cite{Berry1997}. The localization length is then $\xi_1^\textrm{s}\sim l_1^\textrm{s}=1/\sqrt{8\rho_2}$ since the refractive index at resonance is
\begin{equation}
    n^2=1+4i\frac{\rho_2}{k^2}.\label{eq:index}
\end{equation}
This qualitative analysis is confirmed by direct numerical simulations, where the averaged localization length $\xi = \left\langle \xi_m \right\rangle$ and skin depth $\delta = \left\langle \delta_m \right\rangle $ are extracted from exponential fits of the spatial decay of the surface modes: As can be seen in Fig.\ref{fig:2d1d}(a), the localization length of the surface modes scales with the 1D mean-free path $l_1^\textrm{s}$, but scales quite differently from both the 2D mean-free path $l_2$ and from the prediction of the self-consistent theory, which assumes a weak disorder~\cite{Lee1985,Wolfle2010}: $\xi_2^\mathrm{wd}=l_2 \sqrt{\exp(\pi kl_2)-1}$.
Furthermore, even though the skin depth measured in the simulations and that predicted from Eq.\eqref{eq:skindepth} differ by a factor $\sim10$, they present the predicted scaling $\delta_2\sim 1/\sqrt{\rho_2}$. A possible source for this discrepancy is the fact that the skin depth \eqref{eq:skindepth} normally stands for the evanescent wave outside the medium, in addition to the fact that Eq.\eqref{eq:index} for the refractive index has a limited validity for high densities and close to resonance.

As the effective medium theory does not predict any angular confinement (the scattering modes by an homogeneous dielectrics are provided by the spherical harmonics in Mie theory~\cite{vandeHulst1957}), the angular confinement is due to the disorder, i.e., Anderson localization. On the other hand, the difference in the radial and angular decay lengths shows that two distinct confining mechanisms are at play. We therefore attribute our findings to surface Anderson localization of light. 

The presence of surface propagation must be associated to a strong modification of the local density of states (LDOS) close to the surface, which results in an anisotropic random walk of the photons between the scatterers. In particular, the LDOS of an effective-medium theory should capture the preferential emission along the surface, whereas applied to the microscopic model used in this work~\cite{Antezza2013}, it may allow to investigate in more details the differences between the radial and superficial confinements.

\section{Surface localization in 3D\label{sec:3d}}

In three dimensions, localization of light has been more elusive: Experimentally, Anderson localization of light has not yet  been observed in 3D. Even from a theoretical point of view, the existence of a critical density implies that very large number of scatterers are required to overcome finite-size effects in numerical simulations. A consequence of numerical limitations is that localized modes usually present hybrid spatial profiles, with slowly-decaying tails~\cite{Moreira2019, Celardo2017}. 

\begin{figure}[!h]
\centering
\includegraphics[width=0.48\textwidth]{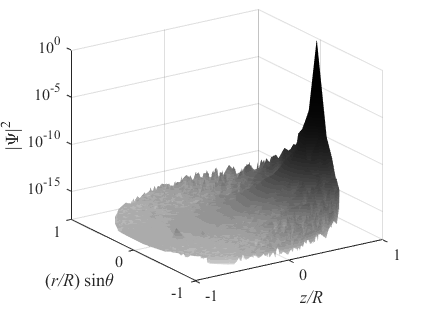} 
\caption{Surface localized mode from the 3D vectorial model~\eqref{eq:3d}. The coordinates refer to spherical ones, with the $z$-axis chosen to cross the mode center of mass. The color code refers to the mode distribution $|\Psi_j|^2$. Simulations realized for a cloud of $N=5000$ scatterers, with a density $\rho=100 k^3$.\label{fig:3dmode} }
\end{figure}

In 3D, for the confinement along the surface, we consider modes with center of mass closer to the surface than $R/20$, and analyze the mode using only the atoms in this layer. An example of such mode is shown in Fig.\ref{fig:3dmode}. The surface localization length is obtained by exponential fitting of the mode profile over the atoms $j$ such that $|R-r_j|\leq R/20$, using the great-circle distance. The surface atoms farthest from the center of mass and that constitute an overall $0.005\%$ of the mode weight are excluded from the fitting~\cite{Moreira2019}.

\begin{figure}[!h]
\centering
\includegraphics[width=0.48\textwidth]{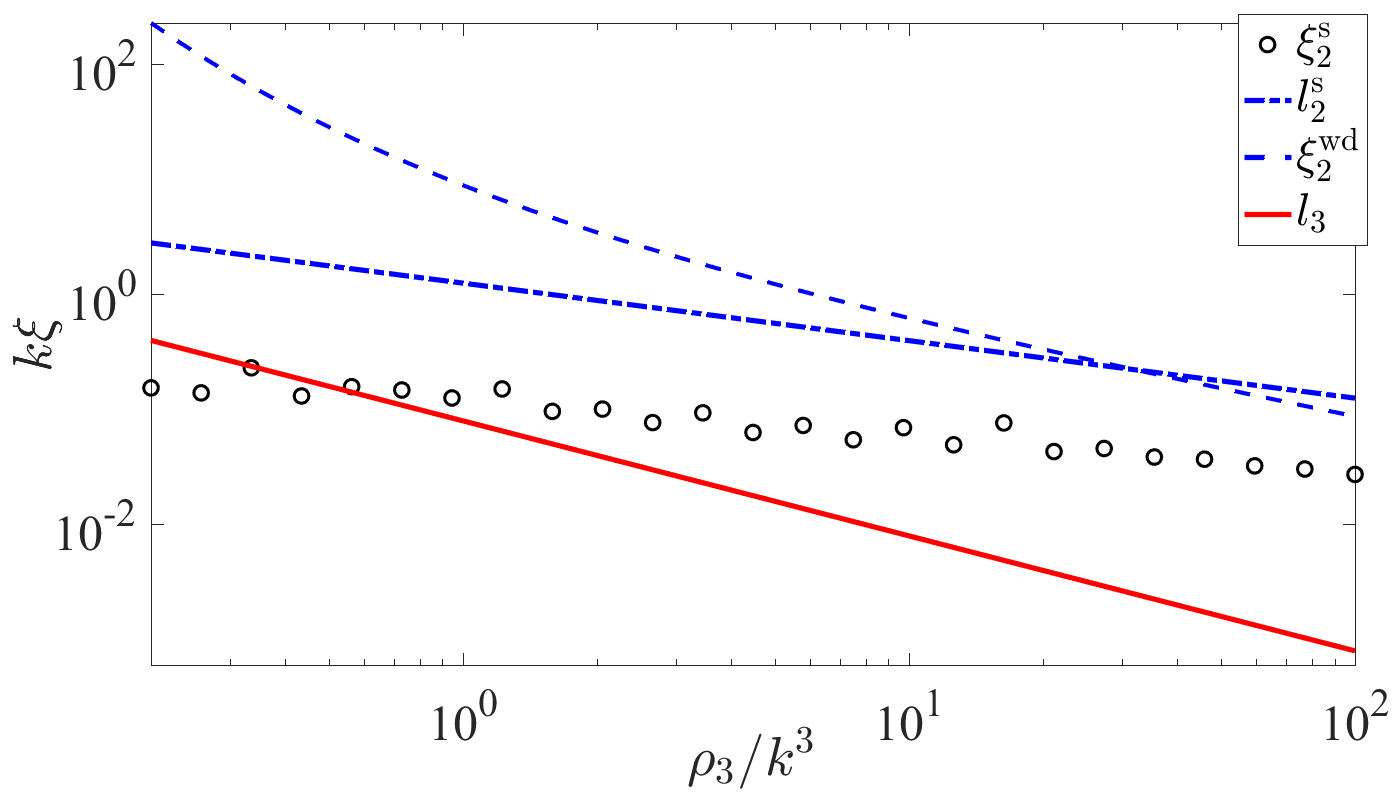} 
\caption{Localization length $\xi_2^\text{s}$ of the surface modes for the 3D vectorial model \eqref{eq:3d}, compared to the 3D mean free path $l_3$ and to the 2D scattering mean free path $l_2^\textrm{s}$ and weak-disorder prediction $\xi_2^\textrm{wd}$ (see main text for details). Simulations realized for a cloud of $N=5000$ scatterers, where $\xi$ is computed using an average over all localized surface modes, and for a single realization for each value of density.\label{fig:3d2d}}
\end{figure}

The obtained $\xi_2^\text{s}$ is presented in Fig.\ref{fig:3d2d}, as a function of the cloud atomic density - obtained from an average over all localized modes for each density. Similarly to the 2D case, the 3D $\xi_2^\text{s}$ scales with the predicted 2D mean-free path along the surface, $l_2^\mathrm{s}$, but scales quite differently from the 3D mean free path $l_3$. We note that $\xi_2^\text{s}$ also scales differently from the 2D self-consistent theory formula~\cite{Lee1985}, yet this prediction assumes a weak disorder, an assumption not verified in our simulations~\cite{Skipetrov2018}. We remark that it was not possible to associate a polarization to the surface modes, when monitoring the orientation of the dipoles. This may be due to the relatively small size of the samples, which can result in a strong curvature of the surface and prevent a well-defined polarization.

\section{Discussion\label{sec:discussion}}

The model used in this paper has been applied extensively to two-level systems.
However, the confinement of light along the surface reported here is similar to
whispering gallery modes, and is thus expected to also
occur with non-resonant dielectrics scatterers. The angular
localization on the other hand is based on the existence of Anderson
localization in dimension $d-1$ for such materials, an effect that has
already been demonstrated for non resonant dielectric media in dimension 1 and 2.
We therefore expect our findings to also hold for non resonant scattering media.

Finally, we note that the edge states discussed in this work are present in both the scalar and vectorial models. In this context, the introduction of a magnetic field introduces two effects that may present an interesting interplay: it splits the scattering channels, whose coupling was preventing localization in 2D and 3D~\cite{Maximo2015,Skipetrov2014}, and it breaks the time-reversal symmetry, thus favoring the emergence of topological insulators~\cite{Hasan2010}. Combining these two effects may thus allow to tune the topological properties of edge states in light scattering systems.

In conclusion, we have shown that although vectorial light scattering systems do not feature a transition to Anderson localization, the confinement of light to the surface in strongly-scattering samples results in disorder-induced localization along the surface. In two dimensions, where large systems can be simulated, a clear distinction between radial and angular confinements can be monitored, and the associated confinement lengths exhibit scalings in agreement with lower-dimensional scattering. In three-dimensional systems, finite-size effects prevent a refined analysis, yet the scaling of the localization length scales as expected from a two-dimensional theory. Considering the size of the samples used for both experimental and theoretical studies of Anderson localization of light, these boundary modes must be accounted for carefully.

\acknowledgments
R. B. and C. E. M. benefited from Grants from São Paulo Research Foundation (FAPESP) (Grants Nos. 2018/01447-2, 2018/15554-5, 2017/13250-6). R. B. and R. K. received support from project CAPES-COFECUB (Ph879-17/CAPES 88887.130197/2017-01). R.B. received support from the National Council for Scientific and Technological Development (CNPq) Grant Nos. 302981/2017-9 and 409946/2018-4. Part of this work was performed in the framework of the European Training Network ColOpt, which is funded by the European Union (EU) Horizon 2020 programme under the Marie Sklodowska-Curie action, grant agreement No. 721465. The Titan X Pascal used for this research was donated by the NVIDIA Corporation.

\bibliography{BiblioCollectiveScattering}

\end{document}